\documentclass[preprint,prl,letterpaper,10pt]{revtex4}%
\usepackage{amsfonts}
\usepackage{amsmath}
\usepackage{amssymb}
\usepackage{graphicx}%
\begin{document}
\preprint{ }
\title[ ]{Influence of Cooper pairing on the inelastic processes in a gas of Fermi atoms.}
\author{V.S.Babichenko }
\affiliation{R.S.C. "Kurchatov Institute", Kurchatov Square 1, Moscow 123182, Russia. }
\author{Yu.Kagan}
\affiliation{R.S.C. "Kurchatov Institute", Kurchatov Square 1, Moscow 123182, Russia. }
\keywords{}
\pacs{}

\begin{abstract}
Correlation properties in ultracold Fermi gas with negative scattering length
and its impact on the three-body recombination is analyzed. We find that
Cooper pairing enhances the recombination rate in contrast to the decrease of
this rate accompanying Bose-Einstein condensation in a Bose gas. This trend is
characteristic for all interval of temperatures T%
$<$%
Tc.

\end{abstract}
\volumeyear{ }
\volumenumber{ }
\issuenumber{ }
\eid{ }
\date{}
\received[Received text]{}

\revised[Revised text]{}

\accepted[Accepted text]{}

\published[Published text]{}

\startpage{0}
\endpage{1}
\maketitle

The investigations of ultracold atomic gases give unique possibilities for
studying quantum correlations and their role in the kinetics and dynamical
properties of macroscopic systems. At early stage of these investigations it
has been predicted that the rate of inelastic processes drastically decreases
with the Bose condensation due to principal reconstruction of quantum
correlations \cite{KSS1}. The local three-particle correlator $K_{3}$,
responsible for the recombination in a dilute gas, reduces with decreasing the
temperature $T<T_{C}$ and the ratio $K_{3}\left(  T\right)  /K_{3}\left(
T_{C}\right)  $ becomes equal to $\sim$1/6 as $T\rightarrow0$. As the gas
parameter $na^{3}$ increases the effect decreases and practically vanishes at
the liquid densities ($a$ is the scattering length).

For the first time, the effect has been observed in work \cite{Wie}. The
authors studied the decay kinetics of a ultracold $^{87}$Rb gas in a trap
under the BEC realization. The experimental results turned out in a
qualitative and quantitative agreement with the theoretical predictions. In
fact, these experimental results were ones of first evidences for the quantum
correlation formation in the process of the BEC kinetics with a limited
lifetime of the system.

The analogous phenomenon takes place in the 2D case at finite temperatures
$T<T_{C}$ when the condensate is absent \cite{KSS2}. This result reflects a
key role of the local correlations which are close to the case of a genuine condensate.

The condensate of Cooper pairs occurs in a two-component atomic Fermi gas with
the attraction between particles at sufficiently low temperatures. It is
interesting to reveal how the Cooper pairing and pair condensate affect the
three-particle recombination rate. Analyzing this problem, we suppose that the
attraction is a result of the Feshbah resonance for s-wave scattering
particles. Herewith, the interaction is supposed to have nonzero value for the
particles of different components only.

The results obtained in the present work lead to a priory not evident
conclusion, namely, Cooper pairing results in enhancing the probability of
three-particle recombination. To make the investigation more transparent, we
consider the low density gas, assuming the following conditions%

\begin{equation}
\mid a\mid k_{F}<<1\text{ \ \ \ and\ \ \ \ }r_{0}k_{F}<<1 \label{1}%
\end{equation}

where $r_{0}$ is the characteristic spatial size of the interparticle interaction.

In the case of negative scattering length $a<0$ the weakly bounded dimers,
typical for $a>0$, are not created. The recombination in this case is
accompanied by the formation of a molecule at a deep energy level. Completely
three-particle character of the relaxation in the case of the negative
scattering length $a<0$, including unitary region, is demonstrated in the work
\cite{Thom}. The formation of the weakly bounded dimers in the case of the
positive scattering length $a>0$ changes the kinetics of the strongly bound
molecule formation essentially \cite{Sal}. The large released energy transfers
to the kinetic energy of the molecule as a whole and to the third particle
involved into the recombination process. Here the picture is analogous for
Bose and Fermi particles and the transition rate is proportional to the
three-particle correlator in the both cases. However, this correlator for
Fermi particles demonstrates the increase with the Cooper pairing in contrast
to its decrease in the BEC case.

2. Considering three-particle recombination in a low density Fermi gas with
the creation of a molecule in a strongly bound state, we present the
Hamiltonian of the system in the form%

\begin{equation}
\widehat{H}=\widehat{H}_{0}+\widehat{H}^{\prime} \label{2}%
\end{equation}

where $\widehat{H}_{0}$ is the Hamiltonian corresponding to elastic processes
and $\widehat{H}^{\prime}$ is the Hamiltonian corresponding to inelastic
recombination processes, respectively,%

\begin{equation}
\widehat{H}^{\prime}=\frac{1}{2}%
{\displaystyle\sum\limits_{\sigma\neq\sigma^{\prime}}}
{\displaystyle\int}
d^{3}r_{1}d^{3}r_{2}d^{3}r_{3}\left\{
\begin{array}
[c]{c}%
\widehat{\psi}_{m}^{+}\left(  \overrightarrow{r}_{1},\overrightarrow{r}%
_{2}\right)  \widehat{\psi}_{\sigma}^{+}\left(  \overrightarrow{r}_{3}\right)
\times\\
\times\widehat{V}\widehat{\psi}_{\sigma}\left(  \overrightarrow{r}_{1}\right)
\widehat{\psi}_{\sigma^{\prime}}\left(  \overrightarrow{r}_{2}\right)
\widehat{\psi}_{\sigma}\left(  \overrightarrow{r}_{3}\right)  +h.c.
\end{array}
\right\}  \label{3}%
\end{equation}

In this expression $\widehat{\psi}_{m}^{+}$\ is the creation operator of a molecule%

\begin{equation}
\widehat{\psi}_{m}^{+}\left(  \overrightarrow{r}_{1},\overrightarrow{r}%
_{2}\right)  =%
{\displaystyle\sum\limits_{\overrightarrow{q}}}
\exp\left(  -i\frac{1}{2}\overrightarrow{q}\left(  \overrightarrow{r}%
_{1}+\overrightarrow{r}_{2}\right)  \right)  \varphi_{m}\left(
\overrightarrow{r}_{1}-\overrightarrow{r}_{2}\right)  \widehat{b}%
_{m\overrightarrow{q}} \label{4}%
\end{equation}

where $\varphi_{m}$ is the wave function of a molecule in its
center-of-inertia frame. The Fermi operators of particle annihilation have the
standard form%

\begin{equation}
\widehat{\psi}_{\sigma}\left(  \overrightarrow{r}\right)  =%
{\displaystyle\sum\limits_{\overrightarrow{k}}}
\widehat{c}_{\overrightarrow{k}\sigma}\exp\left(  i\overrightarrow
{k}\overrightarrow{r}\right)  \label{5}%
\end{equation}

Conserving the pairwise structure of the interaction, we present $\widehat{V}$
in the form $\widehat{V}=U\left(  \overrightarrow{r}_{1}-\overrightarrow
{r}_{2}\right)  +U\left(  \overrightarrow{r}_{3}-\overrightarrow{r}%
_{2}\right)  $. From (4) it follows that $\mid\overrightarrow{r}%
_{1}-\overrightarrow{r}_{2}\mid$\ in (3) has a scale of the molecule size
$r_{\ast}$. The characteristic value of $\mid\overrightarrow{r}_{3}%
-\overrightarrow{r}_{2}\mid$\ in (3) is close to $r_{\ast}$ since in this case
only the third particle should get the momentum compared in the magnitude with
the momentum of the created molecule. Analyzing (3) in the frame which origin
is at the center-of-inertia of the three-particle ensemble, we introduce the variables%

\begin{equation}
\overrightarrow{\rho}=\overrightarrow{r}_{1}-\overrightarrow{r}_{2}\text{,
\ \ \ \ }\overrightarrow{\rho}^{\prime}=\overrightarrow{r}_{3}-\overrightarrow
{r}_{2}\text{\ } \label{6}%
\end{equation}

The bare vertex for a product of three Fermi operators on the right hand-side
with taking (6), (1) and the transposition symmetry into account can be
transformed as%

\begin{align}
&  \exp\left[  i\overrightarrow{k}_{1}\overrightarrow{r}_{1}+i\overrightarrow
{k}_{2}\overrightarrow{r}_{2}+i\overrightarrow{k}_{3}\overrightarrow{r}%
_{3}\right] \label{7}\\
&  \rightarrow\frac{1}{2}\exp\left(  i\overrightarrow{k}_{1}\overrightarrow
{\rho}+i\overrightarrow{k}_{3}\overrightarrow{\rho}^{\prime}\right)  \left[
1-\exp\left(  -i\left(  \overrightarrow{k}_{1}-\overrightarrow{k}_{3}\right)
\left(  \overrightarrow{\rho}-\overrightarrow{\rho}^{\prime}\right)  \right)
\right]  \approx\nonumber\\
&  \approx\frac{i}{2}\exp\left(  i\overrightarrow{k}_{1}\overrightarrow{\rho
}+i\overrightarrow{k}_{3}\overrightarrow{\rho}^{\prime}\right)  \left(
\overrightarrow{k}_{1}-\overrightarrow{k}_{3}\right)  \left(  \overrightarrow
{\rho}-\overrightarrow{\rho}^{\prime}\right) \nonumber
\end{align}

When the energy of a created molecule $E_{m}>>\varepsilon_{F}$,\ the momentum
of the third particle and the momentum of the molecule are practically the
same in the absolute magnitude. Taking into account the relation%

\[
\frac{1}{2}\left(  \overrightarrow{r}_{1}+\overrightarrow{r}_{2}\right)
-\overrightarrow{r}_{3}=\frac{1}{2}\overrightarrow{\rho}+\overrightarrow{\rho
}^{\prime}%
\]

for the vertex in (3), we have%

\begin{align}
&  \Gamma\left(  \overrightarrow{q},\overrightarrow{k}_{1},\overrightarrow
{k}_{2},\overrightarrow{k}_{3}\right) \label{8}\\
&  =\frac{i}{4}%
{\displaystyle\int}
d^{3}\rho d^{3}\rho^{\prime}\left\{
\begin{array}
[c]{c}%
\exp\left[  -i\overrightarrow{q}\left(  \frac{1}{2}\overrightarrow{\rho
}+\overrightarrow{\rho}^{\prime}\right)  +i\left(  \overrightarrow{k}%
_{1}\overrightarrow{\rho}+\overrightarrow{k}_{3}\overrightarrow{\rho}^{\prime
}\right)  \right]  \times\\
\times\left(  \overrightarrow{k}_{1}-\overrightarrow{k}_{3}\right)  \left(
\overrightarrow{\rho}-\overrightarrow{\rho}^{\prime}\right)  \varphi
_{m}\left(  \overrightarrow{\rho}\right)  U\left(  \overrightarrow{\rho
}^{\prime}\right)  +\\
+\left(  \overrightarrow{\rho}\rightleftarrows\overrightarrow{\rho}^{\prime
}\right)
\end{array}
\right\} \nonumber
\end{align}

Considering the Fermi gas with the large magnitude of negative scattering
length $\mid a\mid>>r_{0}$, $r_{\ast}$ and conserving limitation (1), we face
with a sharp increase of vertex $\Gamma$ under conditions of the Feshbah
resonance. In fact, for the singlet pairs a quasi resonance state is realized
in the continuous spectrum for energy $\varepsilon\rightarrow0$, whereas the
real dimer bound state is absent. The solution of the Schroedinger equation
for the two-particle problem demonstrates that, for $\mid\overrightarrow{\rho
}$ $\mid<r_{0}$, the wave function obtains an additional large factor $\mid
a\mid/r_{0}$. The probability amplitude to find simultaneously three fermions
in the volume $\sim r_{\ast}^{3}$ requires a factor $\left(  \mid a\mid
/r_{0}\right)  ^{2}$ as compared with the case of noninteracting gas (c.p.
\cite{Pet}, \cite{BH}, \cite{PHH}). Note that $r_{\ast}\lesssim r_{0}$.
Correspondingly, vertex (8) obtains effectively an additional factor
$\sim\left(  \mid a\mid/r_{0}\right)  ^{2}$.

The integral in (8) can be represented as a sum of two integrals. Each of them
is factorized as a product of two integrals over $\overrightarrow{\rho}$ and
$\overrightarrow{\rho}^{\prime}$. Integrating in every case by parts and
taking into account that $\mid\overrightarrow{k}_{i}$ $\mid<<$ $\mid
\overrightarrow{q}$ $\mid$, we find.%

\begin{equation}
\Gamma=\widetilde{\Gamma}\left(  \overrightarrow{q}\right)  \frac{\left(
\overrightarrow{k}_{1}-\overrightarrow{k}_{3}\right)  \overrightarrow{q}%
}{q^{2}} \label{9}%
\end{equation}

\[
\widetilde{\Gamma}\left(  \overrightarrow{q}\right)  =\xi\frac{i}{2}\left(
\frac{\mid a\mid}{r_{0}}\right)  ^{2}q\left[  \frac{d\varphi_{m}\left(
q/2\right)  }{dq}U\left(  q\right)  +\varphi_{m}\left(  q/2\right)
\frac{dU\left(  q\right)  }{dq}\right]
\]

Here $\xi$ is the numerical coefficient. In all cases we suppose the spherical
symmetry of the functions $U\left(  \overrightarrow{r}\right)  $, $\varphi
_{m}\left(  \overrightarrow{r}\right)  $. As a result, $\widetilde{\Gamma
}\left(  \overrightarrow{q}\right)  $ depends on the absolute value of vector
$\overrightarrow{q}$ alone.

Considering the recombination transitions and, correspondingly, $\widehat
{H}^{\prime}$ as a small perturbation, the number of transitions per unit time
takes the form $\left(  \hbar=1\right)  $%

\begin{equation}
W=2\pi%
{\displaystyle\sum\limits_{i,f}}
\widehat{\rho}_{i}\mid\widehat{H}_{fi}^{\prime}\mid^{2}\delta\left(
E_{f}-E_{i}\right)  =\int dt<\widehat{H}^{\prime}\left(  0\right)  \widehat
{H}^{\prime}\left(  t\right)  > \label{10}%
\end{equation}

\[
\widehat{H}^{\prime}\left(  t\right)  =\exp\left(  i\widehat{H}_{0}t\right)
\widehat{H}^{\prime}\exp\left(  -i\widehat{H}_{0}t\right)
\]

Here the operator $\widehat{\rho}_{i}$ is the equilibrium density matrix
defined by the Hamiltonian $\widehat{H}_{0}$. Summing over index f in (10), we
use the inequality $\varepsilon_{F}<<E_{m}$. Then the block%

\[%
{\displaystyle\sum\limits_{\overrightarrow{q}}}
\mid\widetilde{\Gamma}\left(  q\right)  \mid^{2}\frac{\left(  \left(
\overrightarrow{k}_{1}-\overrightarrow{k}_{3}\right)  \overrightarrow
{q}\right)  }{q^{2}}\frac{\left(  \left(  \overrightarrow{k}_{1}^{\prime
}-\overrightarrow{k}_{3}^{\prime}\right)  \overrightarrow{q}\right)  }{q^{2}}%
{\displaystyle\int\limits_{-\infty}^{\infty}}
dt\exp\left(  -i\left(  E_{m}-\frac{3}{4}\frac{q^{2}}{m}\right)  t\right)
\]

can be singled out in (10). Here the value $\frac{3}{4}\left(  q^{2}/m\right)
$ is the overall kinetic energy of the created molecule and fast atom with the
momenta equal in magnitude and opposite in the direction. The direct
calculation of this expression gives%

\[
\frac{82}{9\pi}q_{\ast}\mid\widetilde{\Gamma}\left(  q_{\ast}\right)  \mid
^{2}\frac{\left(  \left(  \overrightarrow{k}_{1}-\overrightarrow{k}%
_{3}\right)  \left(  \overrightarrow{k}_{1}^{\prime}-\overrightarrow{k}%
_{3}^{\prime}\right)  \right)  }{q_{\ast}^{2}}%
\]

As a result, for the probability (10) we find%

\begin{equation}
W=B%
{\displaystyle\sum\limits_{\substack{\overrightarrow{k}_{i},\overrightarrow
{k}_{i}^{\prime}\\i=1,2,3}}}
\frac{\left(  \left(  \overrightarrow{k}_{1}-\overrightarrow{k}_{3}\right)
\left(  \overrightarrow{k}_{1}^{\prime}-\overrightarrow{k}_{3}^{\prime
}\right)  \right)  }{q_{\ast}^{2}}K^{\left(  3\right)  }\left(
\overrightarrow{k}_{i},\overrightarrow{k}_{i}^{\prime};\sigma,\sigma^{\prime
}\right)  \label{11}%
\end{equation}

where $K^{\left(  3\right)  }$ is the three-particle correlator%

\begin{equation}
K^{\left(  3\right)  }\left(  \overrightarrow{k}_{i},\overrightarrow{k}%
_{i}^{\prime}\right)  =<\widehat{c}_{k_{1}^{\prime}\uparrow}^{+}\left(
0\right)  \widehat{c}_{k_{2}^{\prime}\downarrow}^{+}\left(  0\right)
\widehat{c}_{k_{3}^{\prime}\uparrow}^{+}\left(  0\right)  \widehat{c}%
_{k_{3}\uparrow}\left(  0\right)  \widehat{c}_{k_{2}\downarrow}\left(
0\right)  \widehat{c}_{k_{1}\uparrow}\left(  0\right)  > \label{12}%
\end{equation}

The value of the constant B is determined by the expression%

\begin{equation}
B=\frac{4}{9\pi}mq_{\ast}\mid\widetilde{\Gamma}\left(  q_{\ast}\right)
\mid^{2} \label{13}%
\end{equation}

The characteristic transition time $1/E_{m}$ is small compared with the
characteristic correlation time in (11), and it determines effectively the
correlator (12) for the coincident time moments.

3. We begin with the calculation of the probability W (11) at zero temperature
$T=0$. In the lack of interparticle interaction the correlator $K^{\left(
3\right)  }$ reads \
\[
K^{\left(  3\right)  }\left(  \overrightarrow{k}_{i},\overrightarrow{k}%
_{i}^{\prime}\right)  =n_{\overrightarrow{k}_{1}\uparrow}n_{\overrightarrow
{k}_{2}\downarrow}n_{\overrightarrow{k}_{3}\uparrow}\left(  \delta
_{\overrightarrow{k}_{1};\overrightarrow{k}_{1}^{\prime}}\delta
_{\overrightarrow{k}_{3};\overrightarrow{k}_{3}^{\prime}}-\delta
_{\overrightarrow{k}_{1};\overrightarrow{k}_{3}^{\prime}}\delta
_{\overrightarrow{k}_{3};\overrightarrow{k}_{1}^{\prime}}\right)
\delta_{\overrightarrow{k}_{2};\overrightarrow{k}_{2}^{\prime}}%
\]

Substituting this expression into (11), we find%

\begin{equation}
W_{0}=B\frac{12}{5}\frac{k_{F}^{2}}{q_{\ast}^{2}}n^{3}, \label{14}%
\end{equation}

where n is the particle density with the fixed spin projection. If the
interaction between particles is taken into account, correlator (12) should be
averaged over the changed ground state. We start from the case when the Cooper
pairing is absent and then find the correlator $K^{\left(  3\right)  }$ in
first order in the parameter $\mid a\mid k_{F}<<1$.

The interaction Hamiltonian for low energy particles can be written as%

\[
\widehat{H}_{int}=-%
{\displaystyle\sum\limits_{\substack{\overrightarrow{p}_{1},\overrightarrow
{p}_{2}\\\overrightarrow{p}_{1}^{\prime}\overrightarrow{p}_{2}^{\prime}}}}
V\left(  \overrightarrow{p}\right)  \widehat{c}_{\overrightarrow{k}%
_{1}^{\prime}\uparrow}^{+}\widehat{c}_{\overrightarrow{k}_{2}^{\prime
}\downarrow}^{+}\widehat{c}_{\overrightarrow{k}_{2}\downarrow}\widehat
{c}_{\overrightarrow{k}_{1}\uparrow},
\]

where $\overrightarrow{p}=\overrightarrow{p}_{1}^{\prime}-\overrightarrow
{p}_{1}$ and an additional condition for summing in this expression is
$\overrightarrow{p}_{1}+\overrightarrow{p}_{2}=\overrightarrow{p}_{1}^{\prime
}+\overrightarrow{p}_{2}^{\prime}$. In first order in the interaction the wave
function of the ground state can be represented as
\[
\psi=\psi_{0}+%
{\displaystyle\sum\limits_{S\neq0}}
\frac{<S\mid\widehat{H}_{int}\mid0>}{E_{0}-E_{S}}\psi_{0}^{\left(  S\right)  }%
\]

The direct calculation of the correlator $K^{\left(  3\right)  }$ with taking
this reconstruction into account results in the appearance of four terms. If
the factor $\left(  \overrightarrow{k}_{1}-\overrightarrow{k}_{3}\right)
\left(  \overrightarrow{k}_{1}^{\prime}-\overrightarrow{k}_{3}^{\prime
}\right)  $\ is involved, these terms give the identical summands in Eq. (11),
and the correlator can be written as
\begin{equation}
K_{1}^{\left(  3\right)  }\left(  \overrightarrow{k}_{i},\overrightarrow
{k}_{i}^{\prime}\right)  =-V\left(  \overrightarrow{k}\right)  \left[
\frac{n_{\overrightarrow{k}_{1}^{\prime}\uparrow}n_{\overrightarrow{k}%
_{2}^{\prime}\downarrow}\left(  n_{\overrightarrow{k}_{2}\downarrow
}+n_{\overrightarrow{k}_{1}\uparrow}\right)  }{\varepsilon_{\overrightarrow
{k}_{1}^{\prime}}+\varepsilon_{\overrightarrow{k}_{2}^{\prime}}-\varepsilon
_{\overrightarrow{k}_{1}}-\varepsilon_{\overrightarrow{k}_{2}}}+\frac
{n_{\overrightarrow{k}_{1}^{\prime}\uparrow}n_{\overrightarrow{k}_{2}^{\prime
}\downarrow}}{\varepsilon_{\overrightarrow{k}_{1}^{\prime}}+\varepsilon
_{\overrightarrow{k}_{2}^{\prime}}-\varepsilon_{\overrightarrow{k}_{1}%
}-\varepsilon_{\overrightarrow{k}_{2}}}\right]  n_{\overrightarrow{k}%
_{3}\uparrow}\delta_{\overrightarrow{k}_{3};\overrightarrow{k}_{3}^{\prime}}
\label{15}%
\end{equation}

where $\overrightarrow{k}=\overrightarrow{k}_{2}^{\prime}-\overrightarrow
{k}_{2}$. In this expression the term with a product of four functions
$n_{\overrightarrow{k}_{i}}$ in the numerator gives zero contribution to (11)
due to the symmetry and can be omitted.

We use the usual method of substituting genuine potential $V\left(
\overrightarrow{r}\right)  $ for an effective potential $\overline{V}\left(
\overrightarrow{r}\right)  $ with the same magnitude of the scattering length
$a$ for small energies $\varepsilon\rightarrow0$ and conserve a possibility to
employ the perturbation theory. The potential $\overline{V}\left(
\overrightarrow{r}\right)  $ can be taken as a simple spherical rectangular
well with the depth $\overline{V}_{0}>>\varepsilon_{F}$ and the radius $R_{0}%
$. Applicability of the perturbation theory implies the validity of the
inequality $\varkappa R_{0}<<1$ where $\varkappa=\sqrt{m\overline{V}_{0}}$,
herewith, the bound state in the well is absent. It can easily be revealed
that, for these requirements, the relation $R_{0}=\eta\mid a\mid$ with
$\eta>>1$ takes place as well as combined inequalities $k_{F}\mid a\mid
<<k_{F}R_{0}<<1$. As a consequence, the Fourier component obeys the equality
$\overline{V}\left(  \overrightarrow{k}\longrightarrow0\right)  \approx\left(
4\pi/m\right)  a\equiv g$ and the function $\overline{V}\left(
\overrightarrow{k}\right)  $ begins drastically to decrease for $kR_{0}\sim1$
$\ \left(  k>>k_{F}\right)  $. The last point is important since with the
substitution (15) into (11) and under the integration over momentum
$\overrightarrow{k}$ the divergent of the second term in the brackets in (15)
is eliminated when the dependence $\overline{V}\left(  \overrightarrow
{k}\right)  $ is taken into account. The magnitude of the integral proves to
be $\sim1/R_{0}$ and the ratio $g/R_{0}$ becomes independent of the scattering
length. In fact, the contribution of the corresponding term in the correlator
$K^{\left(  3\right)  }$ determines some renormalization of the probability
$W_{0}$ (14).

In the first term in (15) all the momenta obey inequality $k_{i}\lesssim
k_{F}$ and the effective interaction can be approximated as $\overline
{V}\left(  \overrightarrow{k}\right)  \simeq g$. Substituting $K_{1}^{\left(
3\right)  }$ into (11), we find the probability of the transition $W_{1}$ and,
at the same time, the general expression for W \ in the normal phase in the
linear approximation in interaction%

\begin{equation}
W_{n}=W_{0}+W_{1}=W_{0}\left(  1-\frac{6}{35\pi}\left(  11-2\ln2\right)
k_{F}\mid a\mid\right)  \label{16}%
\end{equation}

4. Consider now the contribution of the Cooper pairing to the formation of the
three-particle correlator (12). For this purpose, averaging over the ground
state, we should include the anomalous averages%

\begin{equation}
K_{S}^{\left(  3\right)  }=-<\widehat{c}_{\overrightarrow{k}_{1}^{\prime
}\uparrow}^{+}\widehat{c}_{-\overrightarrow{k}_{1}^{\prime}\downarrow}%
^{+}><\widehat{c}_{\overrightarrow{k}_{1}\uparrow}\widehat{c}%
_{-\overrightarrow{k}_{1}\downarrow}>n_{\overrightarrow{k}_{3}}\delta
_{\overrightarrow{k}_{3};\overrightarrow{k}_{3}^{\prime}}\delta
_{\overrightarrow{k}_{2}^{\prime};-\overrightarrow{k}_{1}^{\prime}}%
\delta_{\overrightarrow{k}_{2};-\overrightarrow{k}_{1}}+... \label{17}%
\end{equation}

In this expression there are additional three terms resulting from the
commutations of the momenta and giving the identical contribution to (11).

Using the standard method of the u-v transformation for the calculation of the
anomalous averages, see e.g. \cite{LL}, we have at zero temperature%

\[
<\widehat{c}_{\overrightarrow{k}_{1}\uparrow}\widehat{c}_{-\overrightarrow
{k}_{1}\downarrow}>=\frac{\Delta}{E_{k_{1}}}%
\]

Here $E_{k_{1}}=\sqrt{\mid\Delta\mid^{2}+\left(  \frac{k_{1}^{2}}{2m}%
-\mu\right)  ^{2}}$ and $\Delta$ is the Cooper gap in the one-particle
excitation spectrum. Within the framework of the method considered the
integral in the expression obtained is well known%

\[
\int\frac{d^{3}k_{1}}{\left(  2\pi\right)  ^{3}}\frac{\Delta}{E_{k_{1}}}%
=\frac{2\Delta}{g}%
\]

One can see that the result holds for $T\neq0$ if the gap $\Delta$ is the
temperature-dependent quantity $\Delta\left(  T\right)  $. As a result, after
substituting $K_{S}^{\left(  3\right)  }$ (17) into (11), we find directly%

\begin{equation}
W_{S}\left(  T\right)  =W_{0}\frac{9\pi^{2}}{4}\frac{1}{\left(  k_{F}a\right)
^{2}}\left(  \frac{\Delta\left(  T\right)  }{\varepsilon_{F}}\right)  ^{2}
\label{18}%
\end{equation}

From Eq. (18) it follows the main qualitative result: the Cooper pairing or
condensate of Cooper pairs intensifies the three-particle recombination and
thus the relaxation rate of the attractive Fermi gas. In this aspect the
result differs in kind from the BEC case in a dilute atomic Bose gas in which,
on the contrary, the rate of three-particle recombination decreases. Due to
the behaviour of $\Delta$ as the function of the parameter $k_{F}a$ the
three-particle recombination rate rises exponentially when $k_{F}\mid a\mid$
increases, at least up to $k_{F}\mid a\mid\lesssim1$. The exterpolation of the
results obtained from the region $k_{F}\mid a\mid<<1$ to the region $k_{F}\mid
a\mid\lesssim1$ demonstrates the real possibility of the experimental study of
the effect.

The leading temperature correction for $W_{n}$ is connected with $W_{0}$ and
equal to $\left(  \pi^{2}/3\right)  \left(  T/\varepsilon_{F}\right)  ^{2}$.
Since $T_{c}\simeq\Delta\left(  T=0\right)  $, it is readily seen that, for
the interval $T<T_{c}$, the temperature behavior of W is governed by the
dependence $W_{S}\left(  T\right)  $ (18). \


\begin{thebibliography}{9}                                                                                                %


\bibitem {KSS1}Yu. Kagan, B. V. Svistunov, G. V. Shlyapnikov, JETP Lett., 42,
169 (1985).

\bibitem {Wie}E. A. Burt, R. W. Ghrist, C. J. Myatt, E. A. Cornell, and C. E.
Wieman, Phy. Rev. Lett.,79, 337 (1997).

\bibitem {KSS2}Yu. Kagan, B. V. Svistunov, G. V. Shlyapnikov, JETP, 93, 552 (1987).

\bibitem {Thom}X. Du. Y. Zhang, and J. E. Thomas, Phys. Rev. Lett. 102, 250402 (2009).

\bibitem {Sal}D. S. Petrov, C. Salomon, and G. V. Shlyapnikov, Phys. Rev.
Lett. 93, 090404 (2009).

\bibitem {Pet}D. S. Petrov, Phys. Rev. A 67, 010703 (2003).

\bibitem {BH}E. Braaten, H.-W. Hammer, Phys.Rept., 428, 259 (2006). .

\bibitem {PHH}K. Helfrich, H.-W. Hammer, and D. S. Petrov, Phys. Rev. A81,
042715 (2010).

\bibitem {LL}L. D. Landau, E. M. Lifshitz, "Statistical Physics", Part 2,
Pergamon, 1980.
\end{thebibliography}
\end{document}